\documentclass[aps,pre,reprint,superscriptaddress]{revtex4-2}
\usepackage{graphicx}
\usepackage{dcolumn}
\usepackage{bm}
\usepackage{color}
\usepackage{amsmath, amssymb, amsfonts, mathrsfs}
\usepackage{physics}
\usepackage[colorlinks=true, linkcolor=blue, citecolor=blue, urlcolor=blue]{hyperref}
\usepackage{tikz}
\usepackage{pgfplots}
\usepackage{booktabs} 

\usetikzlibrary{arrows.meta, decorations.markings}

\begin{document}

\preprint{APS/123-QED}

\title{Scaling Behaviors of  Work Cumulants in Slow Isothermal Processes}

\author{Ruohan Xu}
\affiliation{School of Physics, Peking University, Beijing 100871, China}

\author{Yanbo Qiao}
\affiliation{School of Physics, Peking University, Beijing 100871, China}

\author{H. T. Quan}
\email{htquan@pku.edu.cn}
\affiliation{School of Physics, Peking University, Beijing 100871, China}
\affiliation{Collaborative Innovation Center of Quantum Matter, Beijing 100871, China}
\affiliation{Frontiers Science Center for Nano-optoelectronics, Peking University, Beijing, 100871, China}

\date{\today}

\begin{abstract}
We study the cumulants of work in a slow isothermal process for gapped systems. Using the Martin-Siggia-Rose-De Dominicis-Janssen (MSRDJ) formalism and the properties of connected correlation functions, we show that in this process, the $n$-th cumulant of work scales as $1/T^{n-1}$ , where $T$ is the time duration.  This result holds generally for arbitrary smooth protocols.  Furthermore, we derive the coefficients of the cumulants from equilibrium properties. These coefficients are found to be relevant to thermodynamic geometric tensors.
\end{abstract}


\maketitle

\section{Introduction}

The work statistics of slowly driven gapped systems has been under extensive research in the field of non-equilibrium thermodynamics \cite{mazonka1999,cohen2003,speckDistributionWorkIsothermal2004,Maes2006,peliti2007,Saha2008,taniguchi2008,Then2008,Engel2009,Saha2011,park2013,Saha_2014,Holubec_2015,Gongzongping2016,Gengli2019,Salazar2020,feiUniversalScalingWork2021,CJF2023}. For example, it is well known that the excess work scales as $1/T$ (where $T$ is the time duration of the driving process) in finite-time thermodynamics, as both theoretical \cite{salamon1980,mazonka1999,Crooks2007,Schmiedl_2008,espo2010,sivakThermodynamicMetricsOptimal2012,mandalAnalysisSlowTransitions2016,Gongzongping2016} and experimental \cite{maExperimentalTest12020} studies reveal. Furthermore, Crooks et al.~\cite{Crooks2007,sivakThermodynamicMetricsOptimal2012} have proved that the scaling coefficients of both the first and the second order cumulants of work can be expressed in terms of equilibrium correlations, and they also pointed out that there is a geometric structure behind the excess work. Specifically, the excess work (directly related to the second cumulant/variance of work \cite{speckDistributionWorkIsothermal2004}) corresponds to the curve length (known as thermodynamic length \cite{Crooks2007,sivakThermodynamicMetricsOptimal2012}) generated by a metric tensor which they named as the thermodynamic metric.  There is also a generalization of the thermodynamic metric \cite{Blaber2020}, where  supra-Stokes tensors are introduced to describe the 3rd order moment of work. Attempts to construct higher-order metrics from raw moments encountered divergent integrals which came from non-connected correlations. This results in an explicitly time-dependent geometry which blurs the static structure in the parameter space.

While many previous studies considered the Gaussian feature of the work distribution which only contains the first and second order cumulants, it should be noticed that this is not the full story. An obvious counter-example would be the case of isolated systems, especially those with phase transitions where the gap is closed and the work distribution can become non-Gaussian \cite{silvaStatisticsWorkDone2008}. In this case, adiabatic results corresponding to $1/T$ scaling will break down~\cite{feiUniversalScalingWork2021,feiWorkStatisticsQuantum2020, zhangWorkStatisticsQuantum2022}.  Furthermore, for gapped systems, non-Gaussian features manifest as non-zero higher-order cumulants in the slow driving regime. Shitara and Ueda have conjectured~\cite{SheLeGeometricApproachNonequilibrium2018} that the higher-order cumulants exhibit a power-law scaling: $\kappa_n \propto (1/T)^{n-1}$, and this is supported by their numerical results and $1/T$-expansion up to the 3rd order. However, they did not fully prove this conjecture. 

Here, we provide a systematic way to calculate the cumulants of work using Martin-Siggia-Rose-De Dominicis-Janssen (MSRDJ) field theory. By using this approach, we are able to prove a generalized version of the conjecture in Ref.~\cite{SheLeGeometricApproachNonequilibrium2018}. Furthermore, we find the coefficients of the work cumulants are deeply related to the generalization of thermodynamic geometry beyond the 3rd order.

\section{Theoretical setup}

Suppose the system is simply described by a Langevin equation:
\begin{equation}
    \partial_t^2 \varphi + \gamma_0 \partial_t \varphi + \partial_\varphi [V_0(\varphi) + V_t(\varphi)] = \eta_t .
\end{equation}
Here $\gamma_0$ is the friction coefficient, the noise $\eta_t$ satisfies $\langle \eta_t \eta_{t'} \rangle = 2D \delta (t-t') $ where $D$ is the diffusion coefficient, $V_0$ is the confining potential, and $V_t$ is the driving potential. For a driving process, we can choose $t \in [0,T]$, and $V_t(\varphi)|_{t=0}=0$. Also, we specify the initial condition to be the equilibrium distribution. To avoid complex operator algebra, we choose to use MSRDJ field theory. Then the Lagrangian would be
\begin{equation}
    \mathcal{L}_t[\varphi,\psi] =  \left[ -\psi \{\partial_t^2 \varphi + \gamma_0 \partial_t \varphi + \partial_\varphi [V_0(\varphi) + V_t(\varphi)]\} + D\psi^2 \right] .
\end{equation}
We define the static Lagrangian:
\begin{equation}
    \mathcal{L}_0[\varphi,\psi] :=  \left[ -\psi \{\partial_t^2 \varphi + \gamma_0 \partial_t \varphi + \partial_\varphi [V_0(\varphi)]\} + D\psi^2 \right] .
\end{equation}
Then $\mathcal{L}_t = \mathcal{L}_0 -  \psi \partial_\varphi [V_t(\varphi)]$. The characteristic function of work in field theory is given by \cite{mallickFieldtheoreticApproachNonequilibrium2011}
\begin{equation}
    \ln \chi(\nu) = \ln \left( \frac{1}{Z} \int \mathcal{D}\varphi \mathcal{D}\psi \exp\left(S + (i\nu)\int_0^T dt \partial_t V_t(\varphi) \right) \right).
\end{equation}

Here $S=\int_{-\infty}^0 dt \mathcal{L}_0 + \int_0^T dt \mathcal{L}_t $. Notice that we extended the time period to ensure the initial condition.

\section{Analysis of the work cumulants}

\subsection{Case 1: linear driving protocol}

We first start from the simplest case with linear driving. In this case we have the following result:
For a confined and gapped system with a small linear driving protocol $V_t =  \frac{t}{T} V_1$, the $n$-th cumulant of the work distribution scales as $\kappa_n \propto \left( \frac{1}{T} \right)^{n-1}$ for large $T$. We will give a detailed derivation in the following.

Using MSRDJ field theory to express the characteristic function of work,
\begin{equation}
    \ln \chi(\nu) = \ln \left( \frac{1}{Z} \int \mathcal{D}\varphi \mathcal{D}\psi \exp\left(S_0 + \int_0^T dt \frac{i\nu}{T} V_1 + \int_0^T dt \frac{t}{T} \mathcal{V} \right) \right) .
\end{equation}
Here $S_0 = \int_{-\infty}^T dt \mathcal{L}_0 $, and $\mathcal{V} := -\psi \partial_\varphi [V_1(\varphi)]$, which is of the same order as $V_1$. Using the perturbation theory to expand in orders of $V_1$, we have:
\begin{widetext}
\begin{equation}
\begin{aligned}
    \ln \chi  &= \sum_{m=0}^\infty \frac{(i\nu)^m}{m! T^m} \sum_{n=0}^\infty \frac{1}{T^n n!} \int_0^T  dt_1 \dots dt_{m+n} t_1 \dots t_n
    \times \langle \mathcal{V}(t_1) \dots \mathcal{V}(t_n) V_1(t_{n+1}) \dots V_1(t_{m+n}) \rangle_{\mathrm{c}} \\
    & = \sum_{m,n=0}^\infty \frac{(i\nu)^m}{m! n!} \int_0^1  ds_1 \dots ds_{m+n} Ts_1 \dots Ts_n 
    \times \langle \mathcal{V}(Ts_1) \dots \mathcal{V}(Ts_n) V_1(Ts_{n+1}) \dots V_1(Ts_{m+n}) \rangle_{\mathrm{c}} 
    .
\end{aligned}
\end{equation}
Because the original system is gapped, the connected correlation function decays exponentially \cite{bertini2004combinatorial}. By defining the maximum temporal separation $\Delta t_{\max} = \max_{i,j}|t_i - t_j|$ and the correlation function
$ G_{mn}(t_1,...t_n; t_{n+1}...t_{n+m})= \langle \mathcal{V}(t_1) \dots V_1(t_{m+n}) \rangle_c $
, we have (by the exponential clustering property of gapped systems)\cite{bertini2004combinatorial, spohnEquilibriumFluctuationsInteracting1986, glimmQuantumPhysics1987, ruelleStatisticalMechanicsRigorous1999}

\begin{equation}
   |G_{mn}|< A_{mn} \exp(-\gamma  \Delta t_{\max}/n).
\end{equation}
Here $\gamma$ is a uniform constant, and $A_{mn}$ is the prefactor bounding the amplitude.  Then  we fix $s_1$ and integrate out other time coordinates,

\begin{equation}
    \ln \chi = \sum_{m,n=0}^\infty \frac{(i\nu)^m T^n }{m! n!} \left\{\int_0^1 ds_1 d(s_2-s_1) ... d(s_n-s_1) d s_{n+1}... ds_{n+m} (s_1^n + \Delta) G_{mn}  \right\} .
\end{equation}

Here the remaining term is defined as $ \Delta  := \Pi_{i=1}^n s_i-s_1^n   $. If we evaluate the integral of this term using a loose bound
$|\Delta| \leq  \sum_{i=2}^n (|s_i-s_1|)  $,
$$ \int_0^1 ds_1 d(s_2-s_1) ... d(s_n-s_1) |\Delta| \times |G_{mn}| \leq \int_0^1 ds_1 d(s_2-s_1) ... d(s_n-s_1) A_{mn} (\sum_{i=2}^n|s_i-s_1| ) \exp(-\gamma T \frac{\sum_{i=2}^n|s_i-s_1|}{n^2}).  $$    
Integrating the relative time factors $(s_i - s_1)$ against the exponential decay $\exp(-\gamma T|s_i - s_1|/n^2)$ gives an additional suppression factor of $1/(\gamma T)$ compared to the leading term. This leads to the following result
\begin{equation}
    \ln \chi = \sum_{m,n=0}^\infty \frac{T}{m!n!} \left(\frac{i\nu}{T}\right)^m  \left\{ \int_0^1 ds s^n C_{mn}(s,T) + \mathcal{O}\left(\frac{1}{\gamma T}\right) \right\} .
\end{equation}
\end{widetext}
where we define
\begin{equation}
C_{mn}(s,T) :=  \int_0^T d t_2 ... dt_{m+n} G_{mn}(sT, t_2,...t_{m+n})  .  
\end{equation} 
Because of the exponential decay property, we can replace the integral bound with $\pm \infty$ with a $\mathcal{O}(e^{-\gamma \min(s,1-s)T})$ error, which can be absorbed into the $\mathcal{O}(\frac{1}{\gamma T})$ term for any finite $s\in (0,1)$. Then $C_{mn}$ is a constant due to the time-translation symmetry of $S_0$. Defining $\mathcal{C}_m:= \sum_n\frac{1}{n!} \int_0^1 ds s^n C_{mn} $, we have
\begin{equation}
    \ln \chi = \sum_{m=0}^\infty \frac{1}{m!} \left(\frac{i\nu}{T}\right)^m T \left( \mathcal{C}_m + \mathcal{O}\left(\frac{1}{\gamma T}\right) \right) .
\end{equation}
Taking the derivative with respect to $\nu$  directly leads to our scaling relation. 

Also, noticing that
\begin{align}
    & \sum_{n=0}^\infty \frac{1}{  n!} s^n C_{mn}  := K_m\left(s\right) \\ & = \int_{-\infty}^{\infty} dt_2 \dots dt_m \langle V_1(0) V_1(t_2) \dots V_1(t_m) \rangle_{\mathrm{c}, s}  
\end{align}
where $\langle \dots \rangle_s$ means the average is taken with the static action $S_0 + s \int dt' \mathcal{V}(t')$, which is exactly the equilibrium action at time $sT$. This is just a special case of thermodynamic geometric tensors. So we can have a more compact result:
\begin{equation}
    \ln \chi(\nu) = \sum_{m=0}^\infty \frac{1}{m!} \left(\frac{i\nu}{T}\right)^m T \left( \int_0^1 ds K_m(s) + \mathcal{O}\left(\frac{1}{\gamma T}\right) \right) .
\end{equation}
This result is quite interesting, since it not only gives the scaling of $\kappa_n$, but also gives the exact coefficient  using \textit{equilibrium} properties:
\begin{equation}
    \kappa_n = \left(\frac{1}{T}\right)^{n-1} \left( \int_0^1 ds K_n(s) + \mathcal{O}\left(\frac{1}{\gamma T}\right) \right) .
\end{equation}

In fact, this result can be proved in a much more general case, as illustrated in the following discussion.

\subsection{Case 2: generic protocol}

For a confined gapped system with $C^{2,1}$ driving $V_t = \sum_i \lambda^i\left(\frac{t}{T}\right) V_i$, for large $T$ we have
\begin{equation}
    \kappa_n = \left(\frac{1}{T}\right)^{n-1} \left( \int_0^1 ds \dot{\lambda}^{\mu_1} \dot{\lambda}^{\mu_2} \dots \dot{\lambda}^{\mu_n} \mathcal{K}_{n(\mu_1 \mu_2 \dots \mu_n)}(s) + \mathcal{O}\left(\frac{1}{\gamma T}\right) \right) .
\end{equation}
Here
\begin{equation}
    \mathcal{K}_{n(\mu_1 \mu_2 \dots \mu_n)}(s) = \int_{-\infty}^{\infty} dt_2 \dots dt_n \langle  {V}_{\mu_1}(0) {V}_{\mu_2}(t_2) \dots {V}_{\mu_n}(t_n) \rangle_{\mathrm{c},s} .
\end{equation}

The derivation is similar to the linear driving case.   Following previous derivation,
\begin{widetext}
\begin{equation}
    \ln \chi = \sum_{m,n=0}^\infty \frac{(i\nu)^mT^{n}}{m! n!} \left\{ \int_0^1 ds_1 \dots ds_{m+n} \prod_{k=1}^n  \lambda^{\mu_k}\left(s_k\right)  \prod_{j=1}^m  \dot{\lambda}^{\mu_{n+j}}\left(s_{n+j}\right) \langle \mathcal{V}_{\mu_1}(s_1T) \dots \mathcal{V}_{\mu_n}(s_nT)  \dots V_{\mu_{m+n}}(s_{m+n}T) \rangle_{\mathrm{c}} \right\} .
\end{equation}
Using $\left |\prod_{k=1}^n  f^{\mu_k}\left(s_k\right) - \prod_{k=1}^n  f^{\mu_k}\left(s_1\right) \right| \leq C \max(|\Delta s|)$ for $C^{2,1}$ functions on $[0,1]^n$, 
\begin{equation}
\begin{aligned}
    \ln \chi = \sum_{m,n=0}^\infty \frac{(i\nu)^mT^{n}}{m! n!} \Bigg\{ \int_0^1 ds_1 \dots ds_{m+n} &  \left( \prod_{k=1}^n\lambda^{\mu_k}\left(s_1\right) + \mathcal{O} (\max_{i=2}^n(|s_i-s_1|)) \right) \left( \prod_{j=1}^m  \dot{\lambda}^{\mu_{n+j}}\left(s_{1}\right) + \mathcal{O}(\max_{i=n+1}^{n+m}(|s_i-s_1|)) \right) \\
    & \times \langle \mathcal{V}_{\mu_1}(Ts_1) \dots \mathcal{V}_{\mu_n}(Ts_n) V_{\mu_{n+1}}(Ts_{n+1}) \dots V_{\mu_{m+n}}(Ts_{m+n}) \rangle_{\mathrm{c}} \Bigg\} .
\end{aligned}
\end{equation}

Again, the integration of the deviations is controlled by the exponential decay of the correlations, and results in a correction of order $\mathcal{O}\left( \frac{1}{\gamma T} \right)$. As before, resummation leads to
\begin{equation}
    \ln \chi(\nu) = \sum_{m=0}^\infty \frac{1}{m!} \left(\frac{i\nu}{T}\right)^m T \left( \int_0^1 ds \dot{\lambda}^{\mu_1} \dots \dot{\lambda}^{\mu_m} \mathcal{K}_{m(\mu_1 \dots \mu_m)}(s) + \mathcal{O}\left(\frac{1}{\gamma T}\right) \right) .
\end{equation}
We thus derived the power-law scaling of $\kappa_n$ conjectured in Ref. \cite{SheLeGeometricApproachNonequilibrium2018} for arbitrary smooth protocols.
\end{widetext}
This result directly shows the geometric structure behind the work cumulants. For the special case of $n=2$, this tensor $\mathcal{K}_{2(\mu\nu)}$ is just the thermodynamic metric introduced by Crooks and Sivak \cite{Crooks2007,sivakThermodynamicMetricsOptimal2012}. For higher orders, the geometry is not Riemannian, but a special case of Finsler geometry as in \cite{Blaber2020}. However, unlike the $C_{\nu_1 \dots \nu_n}$ tensors in  \cite{Blaber2020}, all $ \mathcal{K}_n $ tensors here are fully time-independent, which is more well-defined in geometry. This property comes from the fact that high-order cumulants only capture the non-Gaussian and thus nontrivial part of the work distribution, excluding the non-decaying background of correlations.

It should also be noticed that this proof depends on the finite gap condition, but not on the detailed balance condition. It will break down if the gap is closed and the $\mathcal{O}\left(\frac{1}{\gamma T}\right)$ term is no longer a small value. On the other hand, generalization to non-equilibrium steady state without detailed balance would be possible.

\section{Example: The Breathing Oscillator}
\label{sec: Example}
To explicitly verify the validity of our field-theoretic framework and the generalized geometric tensors, we consider an exactly solvable toy model: the overdamped harmonic oscillator with a time-dependent stiffness. For simplicity, we set the inverse temperature $\beta = 1$ and the friction coefficient $\gamma_0 = 1$. The potential is given by $V_t(x) = \frac{1}{2}\lambda(t)x^2$. We choose an inverse-time driving protocol $\lambda(t) = \frac{1}{1+ct}$ \cite{park2013}, where $c = \frac{1}{T}$ denotes the driving rate.

The generalized force conjugate to the driving parameter is $P:=\partial_\lambda{V_t} = \frac{1}{2}x^2$. In equilibrium at a fixed $\lambda$, the position $x(t)$ follows an Ornstein-Uhlenbeck process, a Gaussian process with the two-point correlation function $C(t) = \langle x(0)x(t) \rangle_{\mathrm{eq}} = \frac{1}{\lambda} e^{-\lambda |t|}$.

According to case 2, the $n$-th cumulant of work is completely determined by the $n$-point connected correlation function of $ P$. Using Wick's theorem, the connected diagram of $n$ squared variables forms a single cycle. Taking into account the permutation of vertices and the internal exchange of legs, the symmetry factor is exactly $\frac{(n-1)!}{2}$. The connected correlator thus factorizes into a convolution ring of two-point functions:
\begin{equation}
    \langle  P(t_1) \dots  P(t_n) \rangle_{\mathrm{c}} = \frac{(n-1)!}{2} \prod_{\mathrm{cycle}} C(t_i, t_{i+1}) .
\end{equation}

To obtain the local geometric tensor $\mathcal{K}_n(\lambda)$, we integrate over the $n-1$ relative time coordinates. This temporal convolution can be evaluated as a 1-loop integral in the Fourier frequency domain:
\begin{equation}
    I_n = \int_{-\infty}^\infty \frac{d\omega}{2\pi} \left( \frac{2}{\omega^2 + \lambda^2} \right)^n = \frac{(2n-3)!!}{(n-1)! \lambda^{2n-1}} .
\end{equation}

Multiplying by the symmetry factor yields the higher-order local metric tensor:
\begin{equation}
    \mathcal{K}_n(\lambda) = \frac{(n-1)!}{2} \cdot I_n = \frac{(2n-3)!!}{2\lambda^{2n-1}} .
\end{equation}

The $n$-th cumulant $\kappa_n$ is obtained by integrating this local tensor along the driving manifold. Using the time derivative of our protocol $\dot{\lambda} = -c \lambda^2$, the integral simplifies to:
\begin{equation}
\begin{aligned}
    \kappa_n &= \int_0^T dt \, (\dot{\lambda})^n \mathcal{K}_n(\lambda(t)) \\
    &= (-c)^n \frac{(2n-3)!!}{2} \int_0^T \frac{1}{1+ct} dt \\
    &= (-1)^n \frac{(2n-3)!!}{2} c^{n-1} \ln(1+cT) .
\end{aligned}
\end{equation}

This analytic scaling yields $\kappa_n \propto \frac{1}{T^{n-1}}$ for all $n \ge 1$. 

To verify this field-theoretic asymptotic result, we compare it with the exact solution of the characteristic function of work $\Psi(t) = \mathbb{E}(e^{qW})$, where $q$ is the counting field. Following the moment-generating framework driven by an Ornstein-Uhlenbeck process \cite{speckWorkDistributionDriven2011}, the generating function can be analytically derived as (see Appendix~\ref{app:exact_solution} for detailed derivations):
\begin{equation}
\begin{aligned}
    \ln(\mathbb{E}(e^{qW})) = -\frac{1}{2} \ln \Bigg[ & \frac{q - m_-}{m_+ - m_-} (1+cT)^{m_+} \\
    & + \frac{m_+ - q}{m_+ - m_-} (1+cT)^{m_-} \Bigg],
\end{aligned}
\end{equation}
where the parameters $m_\pm$ are defined by $m_\pm = [-(c+2) \pm \sqrt{(c+2)^2 + 8qc}]/{2c}$.  

\begin{figure}[htbp]
    \centering
    \includegraphics[width=\linewidth]{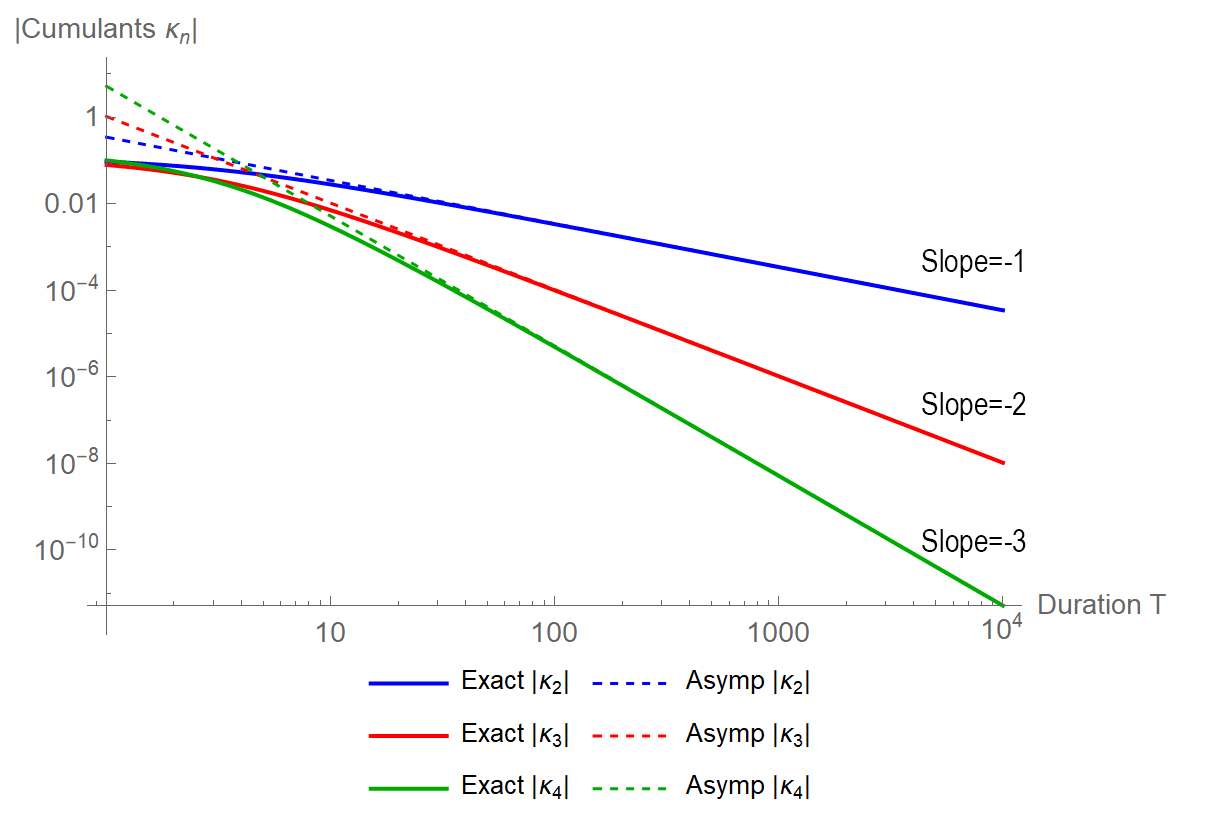} 
    \caption{Comparison between exact (solid) and asymptotic (dashed) results of case 2. The parameters can be found in the first paragraph of Sec.~\ref{sec: Example}}
    \label{fig:comparison}
\end{figure}

Expansion of this solution in the slow-driving limit ($c \to 0$ with $cT$ fixed) reproduces our diagrammatic formula of all cumulants (see Appendix~\ref{app:exact_solution}). This agreement, as visualised in Fig.~\ref{fig:comparison}, demonstrates that the field-theoretic framework captures the scaling behaviors of  work cumulants in  slow driving limit. 

\section{Conclusion}

In conclusion, we have proven the $(1/T)^{n-1}$ scaling behavior of work cumulants conjectured in Ref. \cite{SheLeGeometricApproachNonequilibrium2018} in a slow isothermal process for gapped systems using MSRDJ field theory. This protocol-independent power-law scaling behavior naturally emerges from the dimensional reduction of time integrals, which is guaranteed by the exponential decay of connected correlation functions in gapped systems. 
We also showed that the exact scaling coefficients define a set of higher-order thermodynamic geometric tensors, extending the conventional Riemannian thermodynamic metric \cite{sivakThermodynamicMetricsOptimal2012} to a broader geometric framework.

Finally, we emphasize that this scaling fundamentally relies on the presence of a mass gap. If the system is driven across a continuous phase transition, the correlation time diverges. In such non-adiabatic regimes, the $1/T$ expansion breaks down, leading to the emergence of critical scaling exponents \cite{feiUniversalScalingWork2021,feiWorkStatisticsQuantum2020, zhangWorkStatisticsQuantum2022}.

\begin{acknowledgments}
We acknowledge the support from the National Science
Foundation of China under grants 12375028 and 12521004. We also thank Prof C. Maes for useful discussions.
\end{acknowledgments}

\appendix
\section{Exact Solution for the Breathing Harmonic Oscillator}
\label{app:exact_solution}

In this Appendix, we provide the detailed derivation of the characteristic function of work for the time-dependent breathing oscillator presented in Sec.~\ref{sec: Example}.

Let $\Psi(t) = \mathbb{E}(e^{qW})$ and $\Phi(t) = \langle x^2 e^{qW} \rangle$. According to Ref.~\cite{speckWorkDistributionDriven2011}, their joint time evolution is governed by the following system of ordinary differential equations:
\begin{equation}
    \dot{\Psi} = \frac{q\dot{\lambda}}{2}\Phi, \quad \dot{\Phi} = -2\lambda\Phi + 2\Psi + \frac{3q\dot{\lambda}}{2}\frac{\Phi^2}{\Psi}.
\end{equation}
By introducing the ratio $y(t) = \Phi(t)/\Psi(t)$, the coupled system reduces to a single nonlinear Riccati equation:
\begin{equation}
    \dot{y} = 2 - 2\lambda y + q\dot{\lambda} y^2.
\end{equation}
To linearize this equation, we employ the standard substitution $y(t) = -\frac{1}{q\dot{\lambda}} \frac{\dot{u}}{u}$. Under the protocol $\lambda(t) = \frac{1}{1+ct}$ \cite{park2013}, we change the independent variable to $z = 1+ct$, which transforms the Riccati equation into a second-order linear Euler-Cauchy equation:
\begin{equation}
    z^2 u''(z) + 2\left(1+\frac{1}{c}\right) z u'(z) - \frac{2q}{c} u(z) = 0.
\end{equation}
The general solution of this equation is a linear combination of power functions $u(z) = A z^{m_+} + B z^{m_-}$, where the characteristic roots are given by:
\begin{equation}
    m_\pm = \frac{-(c+2) \pm \sqrt{(c+2)^2 + 8qc}}{2c}.
    \label{eq:m_pm}
\end{equation}
Imposing the initial condition at the equilibrium state $t=0$, we have $y(0) = \langle x^2 \rangle_{\text{eq}} = 1$. Solving for the integration constants $A$ and $B$, and integrating $\dot{\Psi}/\Psi$ back, we arrive at the final expression for the generating function presented in the main text. Expanding Eq.(\ref{eq:m_pm}) at $c \to 0$ and retaining the $\mathcal{O}(c^{n-1})$ order (where $m_-$ term is neglected) yields $\kappa_n = (-1)^n \frac{(2n-3)!!}{2} c^{n-1} \ln(1+cT)$. The remainder contributes $\mathcal{O}(c^n)$ and vanishes.

\bibliographystyle{apsrev4-2}
\bibliography{KZM_Work_Statistics}

\end{document}